\documentclass[runningheads]{llncs}

\usepackage{hyperref}
\usepackage{amsmath}
\usepackage{amssymb}
\usepackage{graphicx}
\usepackage{algorithm}
\usepackage{algorithmic}
\usepackage{subfigure}
\usepackage{threeparttable}

\begin{document}
  \pagestyle{headings}


  \title{Cold-start recommendation through granular association rules}

  \author{Fan Min \and William Zhu}

  \institute{Lab of Granular Computing,\\ Minnan Normal University, Zhangzhou 363000, China\\
  \email{minfanphd@163.com, williamfengzhu@gmail.com}}
  \maketitle

  \begin{abstract}
Recommender systems are popular in e-commerce as they suggest items of interest to users.
Researchers have addressed the cold-start problem where either the user or the item is new.
However, the situation with both new user and new item has seldom been considered.
In this paper, we propose a cold-start recommendation approach to this situation based on granular association rules.
Specifically, we provide a means for describing users and items through information granules, a means for generating association rules between users and items, and a means for recommending items to users using these rules.
Experiments are undertaken on a publicly available dataset MovieLens.
Results indicate that rule sets perform similarly on the training and the testing sets, and the appropriate setting of granule is essential to the application of granular association rules.
    \end{abstract}

  %
  %
  \section{Introduction}\label{section: introduction}
Recommender systems suggest items of interest to users, therefore they serve as an essential part of e-commerce.
To date, hundreds of methods have been proposed.
Collaborative filtering methods \cite{Goldberg1992usingcollaborative,SuXY2009Collaborative} base their recommendations on the historical data of the current user and other users.
Content-based filtering methods \cite{Mooney2000Content,PazzaniM2007Content} involve the attribute information of users and items, and the historical data of the current user.
Hybrid methods \cite{BalabanovicM1997Fab,Cremonesi11Hybrid} integrate these two methods and take advantages of both.
Recently, ensemble methods \cite{Jahrer10Combining,Bothos11Information} are proposed to aggregate predications of base algorithms.

The cold-start problem \cite{Cremonesi11Hybrid,ScheinA2002ColdStart} is difficult in recommender systems.
Researchers have addressed the cold-start problem where either the user or the item is new.
The new item problem \cite{Cremonesi11Hybrid} is more often addressed, and the new user problem is symmetric to it \cite{ScheinA2002ColdStart}.
Naturally, content-based filtering methods \cite{Mooney2000Content,PazzaniM2007Content} are appropriate for these problems.
However, the situation with both new user and new item has seldom been considered.
Since the historical data of the current user and item are unknown, the problem for this situation is more challenging.

In this paper, we propose a cold-start recommendation approach for the new situation based on granular association rules.
This approach is also applicable to existing situations.
First, we provide a means for describing users and items through information granules.
Examples of information granules include ``male students," ``thriller movies," and ``adventure movies released in 1990s".
Then we provide a means for generating association rules between users and items.
An example granular association rule might be ``60\% male students rate 40\% drama movies released in 1990s; 20\%
users are male students and 5\% movies are drama released in 1990s."
Here 60\%, 40\%, 20\%, and 5\% are the source coverage, the target coverage, the source confidence, and the target confidence, respectively.
To obtain strong and meaningful rules, we need to specify thresholds for all four measures.
Finally, we provide a means for recommending items to users using these rules.
The quality of the recommender is evaluated mainly by the recommendation accuracy.

There are already some works on granular association rules.
The concept is defined in \cite{MinHuZhu12GranularFour,MinHuZhu13GranularTwo} with a number of algorithms computing all granular association rules meeting thresholds of four measures.
A more efficient algorithm which takes advantage of rough set theory is presented in \cite{MinZhu12Parametric}.
Two discretization approaches are studied in \cite{HeMinZhu13AComparison} for numeric data.
Multi-value data are then addressed in \cite{MinZhu13Multivalued} to obtain positive rules and discard negative rules such as ``male students rate movies that are \emph{not} comedy."
This is because negative rules are uninteresting in such applications and they overwhelm positive ones.

The main contribution of this paper will be the validation of granular association rules.
In other words, we train and test granular association rules to study their performance.
This work is an important step toward the application of granular association rules \cite{MinHuZhu12GranularFour,MinHuZhu13GranularTwo} as well as granular computing \cite{Lin98Granular,ZhuWang03Reduction,Zhu07Basic,YaoVasilakosPedrycz2013Granular,YaoDeng2013Paradigm,MinHeQianZhu11Test}.
Experiments are undertaken on the MovieLens dataset \cite{movielens} using our open source software Grale \cite{Grale}.
Results indicate that rule sets perform similarly on the training and the testing sets.
More importantly, the appropriate setting of granule is essential to the application of granular association rules.

  %
  %
  \section{Preliminaries}\label{section: preliminaries}
In this section, we present some preliminary knowledge, especially granular association rules.

  %
  %
  \subsection{Many-to-many entity-relationship systems}\label{subsection: scaled-positive}
First we revisit the definitions of information systems, binary relations and many-to-many entity relationship systems \cite{MinHuZhu12GranularFour}.
\begin{definition}\label{definition: ins}
$S = (U, A)$ is an information system, where $U = \{x_1, x_2, \dots, x_n\}$ is the set of all objects, $A = \{a_1, a_2, \dots, a_m\}$ is the set of all attributes, and $a_j(x_i)$ is the value of $x_i$ on attribute $a_j$ for $i \in [1..n]$ and $j \in [1..m]$.
\end{definition}

Two information systems are listed in Tables \ref{subtable: user} and \ref{subtable: movie}, respectively.
In Table \ref{subtable: movie}, 1 indicates \emph{true}, while 0 indicates \emph{false}.

\begin{definition}\label{definition: binary-relation}
Let $U = \{x_1, x_2, \dots, x_n\}$ and $V = \{y_1, y_2, \dots, y_k\}$ be two sets of objects.
Any $R \subseteq U \times V$ is a binary relation from $U$ to $V$.
\end{definition}

An example of binary relation is given by Table \ref{subtable: rates}, where $U$ is the set of users as indicated by Table \ref{subtable: user}, and $V$ is the set of movies as indicated by Table \ref{subtable: movie}.
A binary relation can be viewed as an information system.
However, in order to save storage space, it is more often stored in the database as a table with two foreign keys.

\begin{definition}\label{definition: m-m-er}
A many-to-many entity-relationship system (MMER) is a 5-tuple $ES = (U, A, V, B, R)$, where $(U, A)$ and $(V, B)$ are two information systems, and $R \subseteq U \times V$ is a binary relation from $U$ to $V$.
\end{definition}
An example of MMER is given by Table \ref{table: mmer}.

\begin{table}[tb]\caption{A many-to-many entity-relationship system}\label{table: mmer}
\centering
\setlength{\tabcolsep}{19pt}
\subtable[User]{
\begin{tabular*}{12cm}{@{\extracolsep{\fill}}ccccc}
\hline
User-id     & Age            & Gender &  Occupation \\
\hline
1           & $[18, 24]$     & M      &  technician \\
2           & $[50, 55]$     & F      &  other      \\
3           & $[18, 24]$     & M      &  writer     \\
\dots       \\
943         & $[18, 24]$     & M      &  student    \\
\hline
\end{tabular*}
\label{subtable: user}
}
\qquad
\setlength{\tabcolsep}{2pt}
\subtable[Movie]{
\begin{tabular*}{12cm}{@{\extracolsep{\fill}}cccccccccc}
\hline
Movie-id       & Release-decade &  Action & Adventure & Animation & \dots & Western \\
\hline
1        & 1990s          &  0      & 0         & 0         & \dots & 0\\
2        & 1980s          &  0      & 1         & 1         & \dots & 0\\
3        & 1990s          &  0      & 0         & 0         & \dots & 0\\
\dots    & \\
1,682    & 1960s          &  0      & 0         & 0         & \dots & 0\\
\hline
\end{tabular*}
\label{subtable: movie}
}
\qquad
\setlength{\tabcolsep}{10pt}
\subtable[Rates]{
\begin{tabular*}{12cm}{@{\extracolsep{\fill}}cccccccc}
\hline
User-id$\diagdown$ Movie-id &  1     &  2     &  3    & 4      & 5      & \dots & 1,682\\
\hline
1      &  0      &  1      &  0     &  1      & 0     & \dots  & 0\\
2      &  1      &  0      &  0     &  1      & 0     & \dots  & 1\\
3      &  0      &  0      &  0     &  0      & 1     & \dots  & 1\\
\dots  & \\
943    &  0      &  0      &  1     &  1      & 0     & \dots  & 1\\
\hline
\end{tabular*}
\label{subtable: rates}
}
\end{table}

  %
  %
  \subsection{Information granules}\label{subsection: granules}
Now we employ information granules \cite{MinHuZhu13GranularTwo,YaoDeng2013Paradigm} to describe users and items.
In an information system, any $A' \subseteq A$ induces an equivalence relation \cite{Pawlak82Rough,SkowronStepaniuk94Approximation}
\begin{equation}\label{equation: equivalence-relation}
E_{A'} = \{(x, y) \in U \times U| \forall a \in A', a(x) = a(y)\},
\end{equation}
and partitions $U$ into a number of disjoint subsets called \emph{blocks} or \emph{granules}.
The granule containing $x \in U$ is
\begin{equation}\label{equation: block-contain-x}
E_{A'}(x) = \{y \in U| \forall a \in A', a(y) = a(x)\}.
\end{equation}

The following definition was employed by Yao and Deng \cite{YaoDeng2013Paradigm}.
\begin{definition}\label{definition: granule}
A granule is a triple
\begin{equation}\label{equation: granule}
G = (g, i(g), e(g)),
\end{equation}
where $g$ is the name assigned to the granule, $i(g)$ is a representation of the granule,
and $e(g)$ is a set of objects that are instances of the granule.
\end{definition}

According to Equation (\ref{equation: block-contain-x}), $(A', x)$ determines a granule in an information system.
Hence $g = g(A', x)$ is a natural name to the granule.
$i(g)$ can be formalized as the conjunction of respective attribute-value pairs, i.e.,
\begin{equation}\label{equation: intension-granule}
i(g(A', x)) =  \bigwedge_{a \in A'}\langle a: a(x) \rangle.
\end{equation}
$e(g)$ is given by
\begin{equation}\label{equation: extension-granule}
e(g(A', x)) = E_{A'}(x).
\end{equation}

The \emph{support} of the granule is the size of $e(g)$ divided by the size of the universe, namely,
\begin{equation}\label{equation: support-granule}
supp(g(A', x)) = supp(\bigwedge_{a \in A'}\langle a: a(x) \rangle) = supp(E_{A'}(x)) = \frac{|E_{A'}(x)|}{|U|}.
\end{equation}

Let $x \in U$ and $A'' \subset A' \subseteq A$, we have
\begin{equation}\label{equation: finer-granule}
e(g(A', x)) \subseteq e(g(A'', x)).
\end{equation}
Consequently, we say that $g(A', x)$ is \emph{finer} than $g(A'', x)$, or $g(A'', x)$ is \emph{coarser} than $g(A', x)$.

  %
  %
  \subsection{Granular association rules}\label{subsection: grarule-rules}
Now we discuss the means for connecting users and items.
A \emph{granular association rule} \cite{MinHuZhu12GranularFour,MinHuZhu13GranularTwo} is an implication of the form
\begin{equation}\label{equation: granular-association}
(GR): \bigwedge_{a \in A'}\langle a: a(x) \rangle \Rightarrow \bigwedge_{b \in B'}\langle b: b(y) \rangle,
\end{equation}
where $A' \subseteq A$ and $B' \subseteq B$.

Before defining evaluation measures, let us look at an example of granular association rule ``60\% male students rate 40\% drama movies released in 1990s; 20\% users are male students and 5\% movies are drama released in 1990s."
Here 60\%, 40\%, 20\%, and 5\% are the source coverage, the target coverage, the source confidence, and the target confidence, respectively.

According to Equations (\ref{equation: intension-granule}), (\ref{equation: extension-granule}) and (\ref{equation: granular-association}), the set of objects meeting the left-hand side of the granular association rule is
\begin{equation}\label{equation: left-granular-rule}
LH(GR) = E_{A'}(x),
\end{equation}
while the set of objects meeting the right-hand side of the granular association rule is
\begin{equation}\label{equation: right-granular-rule}
RH(GR) = E_{B'}(y).
\end{equation}

The \emph{source coverage} of $GR$ is
\begin{equation}\label{equation: source-coverage}
scov(GR) = |LH(GR)|/ |U|;
\end{equation}
while the \emph{target coverage} of $GR$ is
\begin{equation}\label{equation: target-coverage}
tcov(GR) = |RH(GR)| / |V|.
\end{equation}

There is a tradeoff between the source confidence and the target confidence.
Consequently, neither of them can be obtained directly from the rule.
To compute any one of them, we need to specify the threshold of the other.
In our computation, we adopt the following approach.
Let $tc$ be the target confidence threshold.
The \emph{source confidence} of the rule is
\begin{equation}\label{equation: source-confidence}
sconf(GR, tc)
= \frac{|\{x \in LH(GR) | \frac{|R(x) \cap RH(GR)|}{|RH(GR)|} \geq tc\}|}{|LH(GR)|}.
\end{equation}

Naturally, the source coverage and the target coverage indicate the generality of a rule, while the source confidence and the target confidence indicate the strength of a rule.

  %
  %
  \section{Training and testing granular association rules}\label{section: train-test}
The set of all possible granular association rules might be very large.
Therefore we would like to train a relatively small rule set.
The key issue is: which rules should be included in this set for recommendation?
One approach is to specify thresholds of four measures mentioned in Section \ref{subsection: grarule-rules}.
All rules satisfying these thresholds are generated to build the recommender.

\begin{problem}\label{problem: partial-rule}
The granular association rule mining problem.

\textbf{Input:} An $ES = (U, A, V, B, R)$, a minimal source coverage threshold $ms$, a minimal target coverage threshold $mt$, a minimal source confidence threshold $sc$, and a minimal target confidence threshold $tc$.

\textbf{Output:} A rule set where each rule satisfying $scov(GR) \geq ms$, $tcov(GR) \geq mt$, $sconf(GR) \geq sc$, and $tconf(GR) \geq tc$.
\end{problem}

Algorithm \ref{algorithm: sandwich} is a straightforward approach to obtaining this rule set.
It is quite similar to the algorithm presented in \cite{MinHuZhu13GranularTwo}.
The difference lies in that the new algorithm stores the rule set in the memory instead of outputting it directly.
Another algorithm presented in \cite{MinZhu12Parametric} can be also revised for this problem to improve efficiency.
Since the focus of the paper is the effectiveness of the algorithm, we will not discuss it here.
\begin{algorithm}[tb!]\caption{A sandwich algorithm for granular association rule mining}\label{algorithm: sandwich}
  \textbf{Input}: $ES = (U, A, V, B, R)$, $ms$, $mt$, $sc$, $tc$.\\
  \textbf{Output}: A rule set.\\
  \textbf{Method}: sandwich.\\
  \begin{algorithmic}[1]
    \STATE $SG(ms) = \{(A', x) \in 2^A \times U| \frac{|E_{A'}(x)|}{|U|} \geq ms\}$;
    \STATE $TG(mt) = \{(B', y) \in 2^B \times V| \frac{|E_{B'}(y)|}{|V|} \geq mt\}$;
    \FOR {each $g \in SG(ms)$}
      \FOR {each $g' \in TG(mt)$}
        \STATE $GR = (i(g) \Rightarrow i(g'))$;
        \IF {$sconf(GR, tc) \geq sc$}
          \STATE add $GR$ to the rule set;
        \ENDIF
      \ENDFOR
    \ENDFOR
  \end{algorithmic}
\end{algorithm}

For each user, we recommend items of interest using the rule set.
All rules that match the user are fired for recommendation.
Hence some users may have many recommendations, and some may have very few.
The performance of the recommender is evaluated mainly by the recommendation accuracy.
Formally, let the number of recommended items be $M$, and the number of success recommendations be $N$, the accuracy is $N / M$.

We will compare five training and testing scenarios.
\begin{enumerate}
\item{Random recommendation. There is no training stage. An item is randomly recommended to a user. This is a baseline approach since a recommender which is not significantly better than a random one is simply useless.}
\item{Test the training set. It is interesting to test the rule set on the training set.
    Since only attribute values can be employed, the performance of the rule set may not be high.}
\item{Divide the user set into the training and testing set. This scenario corresponds to the new user cold-start problem.}
\item{Divide the item set into the training and testing set. This scenario corresponds to the new item cold-start problem.}
\item{Divide both the user and the item sets. In this scenario, both users and items are new. Hence the problem is very challenging.}
\end{enumerate}

  %
  %
  \section{Experiments}\label{section: experiments}
In this section, we try to answer the following questions through experimentation.
\begin{enumerate}
\item{How does the recommender perform for the new user, new item, and both new cold-start problems?}
\item{How does the number of recommendations change for different thresholds?}
\item{How does the performance of the recommender vary for the training/testing sets?}
\item{How does the performance of the recommender change for different threshold settings?}
\end{enumerate}

  %
  %
  \subsection{Dataset}\label{subsection: dataset}
We tested granular association rules on the MovieLens \cite{movielens} which is widely used in recommender systems (see, e.g., \cite{Cremonesi11Hybrid,ScheinA2002ColdStart}).
The database schema is as follows.
\begin{enumerate}
\item[$\bullet$]{User (\underline{userID}, age, gender, occupation)}
\item[$\bullet$]{Movie (\underline{movieID}, release-year, genre)}
\item[$\bullet$]{Rates (\underline{userID, movieID})}
\end{enumerate}
We use the version with 943 users and 1,682 movies.
The data are preprocessed to cope with Definition \ref{definition: m-m-er} as follows.
The original Rate relation contains the rating of movies with 5 scales, while we only consider whether or not a user
has rated a movie.
The user age is discretized to 5 intervals as follows: [7,22), [22,27), [27,31), [31,39), [39,48) and [48,73].
The release year is discretized to 5 intervals as follows: [1922,1980), [1980,1993), [1993,1994), [1994,1995),[1995,1996), [1996,1997) and [1997,1998].
The genre is a multi-valued attribute. Therefore we scale it to 18 boolean
attributes and deal with it using the approach proposed in \cite{MinZhu13Multivalued}.

  %
  %
  \subsection{Results}\label{subsection: results}
We undertake three sets of experiments to answer the questions raised at the beginning of the section one by one.
The settings are as follows: the training set percentage is 60\%, and $sc = tc = 0.3$.
Each experiment is repeated 20 times with different sampling of training and testing sets, and the average accuracy is computed.

\begin{figure}[h]
    \centering
        \includegraphics[width=3.5in]{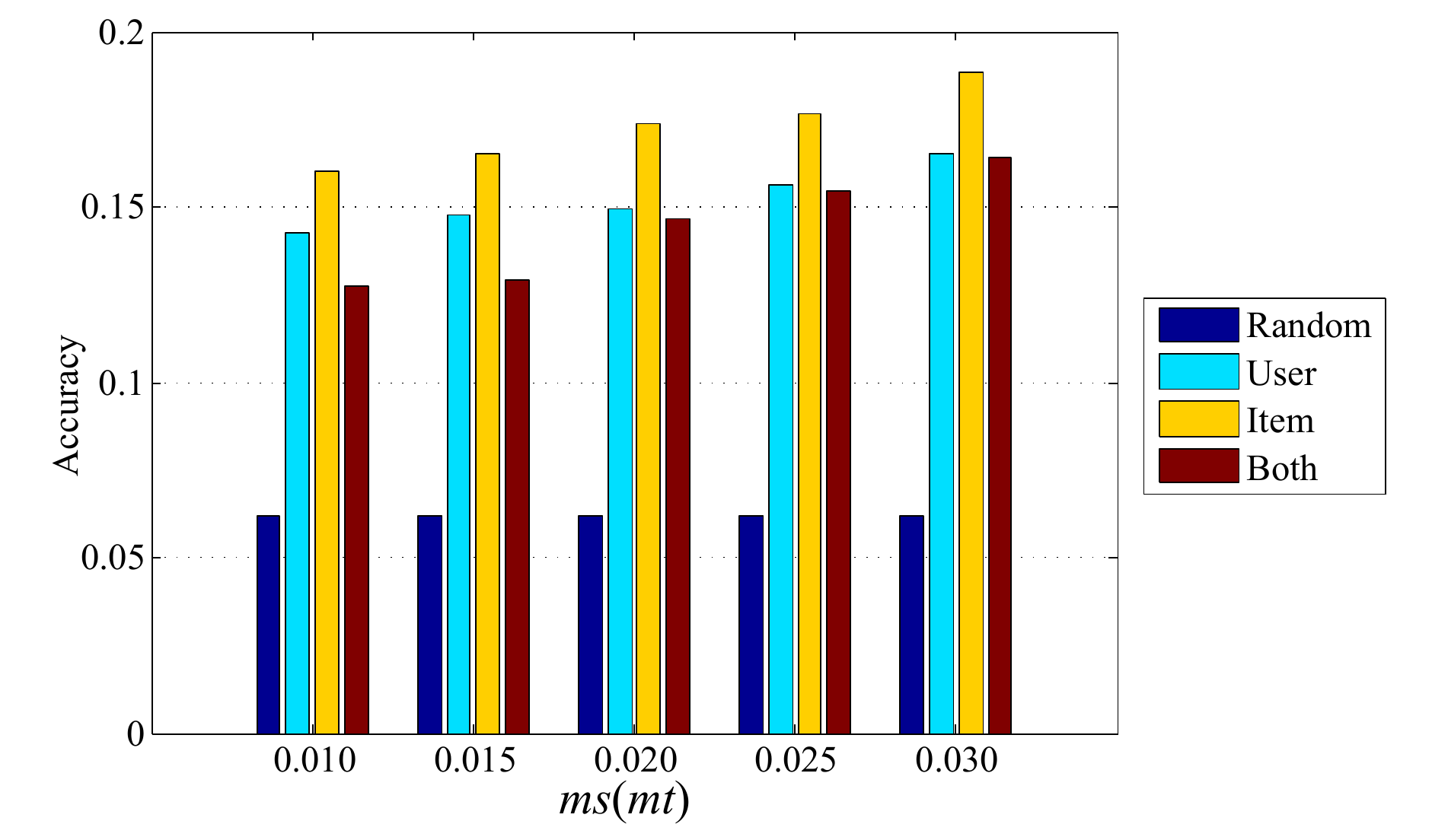}
\caption{Comparison of different scenarios}
\label{figure: scenarios}
\end{figure}

Fig. \ref{figure: scenarios} shows accuracy of the recommender on the new user, new item, and both new scenarios.
The random recommender, which has an accuracy close to 0.062, is also illustrated for comparison.
The result indicates that the new item problem is easier and respective recommendations are more meaningful.
The both new scenario is the hardest since the least information is available.

\begin{figure}[h]
    \centering
        \includegraphics[width=3in]{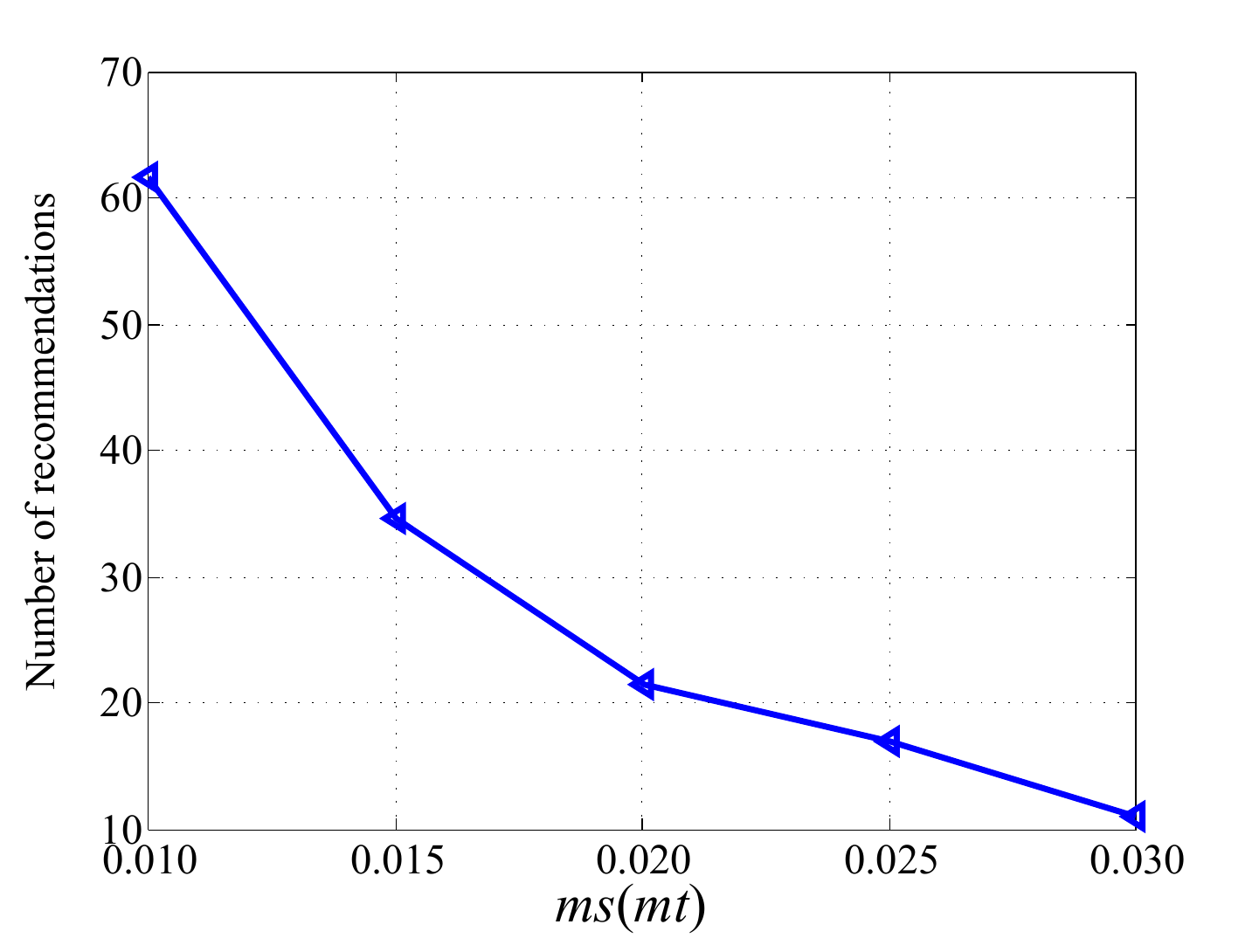}
\caption{Comparison of the number of recommendations}
\label{figure: recommendations}
\end{figure}
Fig. \ref{figure: recommendations} indicates that the number of recommendations decrease dramatically with the increase of $ms$ and $mt$.

\begin{figure}[h]
    \centering
        \includegraphics[width=3.5in]{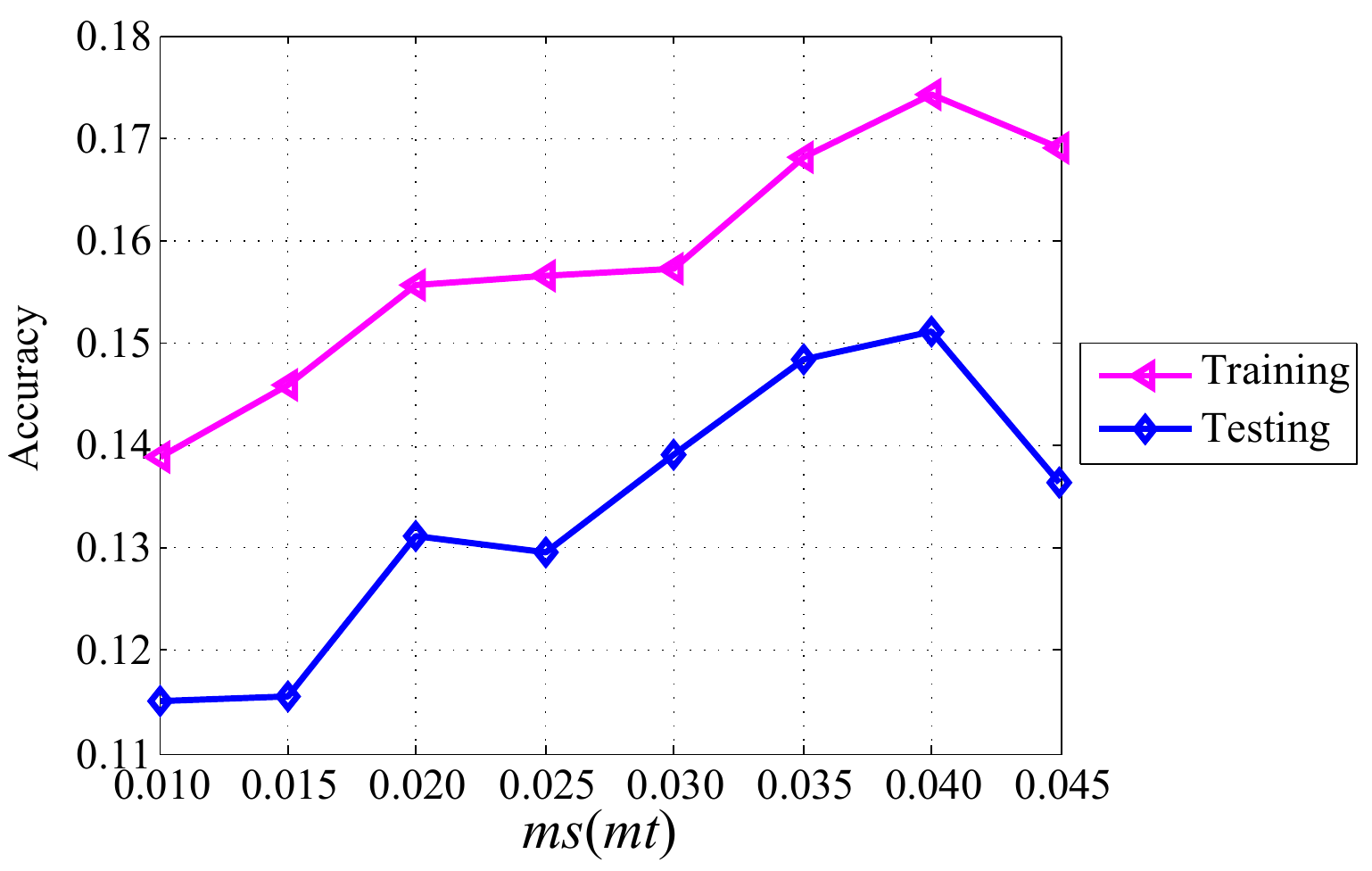}
\caption{Comparison of different granules}
\label{figure: granules}
\end{figure}

Fig. \ref{figure: granules} indicates with the change of $ms$ and $mt$, the recommender performs different.
The following observations are especially interesting from this figure.
\begin{enumerate}
\item When $ms$ and $mt$ are appropriately set (0.04), the recommender has the maximal accuracy.
    With the increase or decrease of these thresholds, the performance of the recommender decreases rapidly.
    This is the most important observation from the experiment.
\item The performance of the recommender does not vary much on the training and the testing sets.
    This phenomenon indicates that the recommender is stable.
\item The recommender achieves the best performance for both the training and the testing set with the same settings.
    This is somehow surprising since the recommender does not incur over-fitting.
\end{enumerate}

  %
  %
  \section{Conclusions}\label{section: conclusion}
In this paper, granular association rules are trained to build cold-start recommender.
We specify different granules in terms of source coverage and target coverage to obtain different recommenders.
Experimental results indicate that the appropriate selection of the granule are essential to the performance of the recommender.
In the future we will design algorithms for effective granule selection.

  %
  %
  \section*{Acknowledgements}\label{section: acknowledgements}
This work is in part supported by National Science Foundation of China under Grant No. 61170128, Fujian Province Foundation of Higher Education under Grant No. JK2012028, and the Natural Science Foundation of Fujian Province, China under Grant No. 2012J01294, and State key laboratory of management and control for complex systems open project under Grant No. 20110106.

  %
  %

\begin{thebibliography}{10}

\bibitem{Goldberg1992usingcollaborative}
Goldberg, D., Nichols, D., Oki, B.M., Terry, D.:
\newblock Using collaborative filtering to weave an information tapestry.
\newblock Communications of the ACM \textbf{35} (1992)  61--70

\bibitem{SuXY2009Collaborative}
Su, X.Y., Khoshgoftaar, T.M.:
\newblock A survey of collaborative filtering techinques.
\newblock Advances in Artificial Intelligence \textbf{2009} (August 2009)
  1--19

\bibitem{Mooney2000Content}
Mooney, R.J., Roy, L.:
\newblock Content-based book recommending using learning for text
  categorization.
\newblock In: Proceedings of the fifth ACM conference on Digital libraries.
  (2000)  195--204

\bibitem{PazzaniM2007Content}
Pazzani, M., Billsus, D.:
\newblock Content-based recommendation systems.
\newblock In: The Adaptive Web. Volume 4321.
\newblock Springer (2007)  325--341

\bibitem{BalabanovicM1997Fab}
Balabanovi\'{c}, M., Shoham, Y.:
\newblock Fab: content-based, collaborative recommendation.
\newblock Communication of ACM \textbf{40}(3) (March 1997)  66--72

\bibitem{Cremonesi11Hybrid}
Cremonesi, P., Turrin, R., Airoldi, F.:
\newblock Hybrid algorithms for recommending new items.
\newblock In: Proceedings of the 2nd International Workshop on Information
  Heterogeneity and Fusion in Recommender Systems. (2011)  33--40

\bibitem{Jahrer10Combining}
Jahrer, M., T\"{o}cher, A., Legenstein, R.:
\newblock Combining predictions for accurate recommender systems.
\newblock In: Proceedings of the 16th ACM SIGKDD. (2010)  693--702

\bibitem{Bothos11Information}
Bothos, E., Christidis, K., Apostolou, D.:
\newblock Information market based recommender systems fusion.
\newblock In: Proceedings of the 2nd International Workshop on Information
  Heterogeneity and Fusion in Recommender Systems. (2011)  1--8

\bibitem{ScheinA2002ColdStart}
Schein, A.I., Popescul, A., Ungar, L.H., Pennock, D.M.:
\newblock Methods and metrics for cold-start recommendations.
\newblock In: SIGIR '02. (2002)  253--260

\bibitem{MinHuZhu12GranularFour}
Min, F., Hu, Q.H., Zhu, W.:
\newblock Granular association rules with four subtypes.
\newblock In: Proceedings of the {IEEE} International Conference on Granular
  Computing. (2012)  432--437

\bibitem{MinHuZhu13GranularTwo}
Min, F., Hu, Q., Zhu, W.:
\newblock Granular association rules on two universes with four measures.
\newblock submitted to Information Sciences (2013)

\bibitem{MinZhu12Parametric}
Min, F., Zhu, W.:
\newblock Granular association rule mining through parametric rough sets.
\newblock In: Proceedings of the 2012 International Conference on Brain
  Informatics. Volume 7670 of LNCS. (2012)  320--331

\bibitem{HeMinZhu13AComparison}
He, X., Min, F., Zhu, W.:
\newblock A comparison of two discretization approches in granular association
  rule mining.
\newblock In: Proceedings of the 2013 Canadian Conference on Electrical and
  Computer Engineering (accepted). (2013)

\bibitem{MinZhu13Multivalued}
Min, F., Zhu, W.:
\newblock Granular association rules for multi-valued data.
\newblock In: Proceedings of the 2013 Canadian Conference on Electrical and
  Computer Engineering (accepted). (2013)

\bibitem{Lin98Granular}
Lin, T.Y.:
\newblock Granular computing on binary relations {I}: Data mining and
  neighborhood systems.
\newblock In: Rough Sets in Knowledge Discovery. (1998)  107--121

\bibitem{ZhuWang03Reduction}
Zhu, W., Wang, F.:
\newblock Reduction and axiomization of covering generalized rough sets.
\newblock Information Sciences \textbf{152}(1) (2003)  217--230

\bibitem{Zhu07Basic}
Zhu, W.:
\newblock Basic concepts in covering-based rough sets.
\newblock In: Proceedings of the International Conference on Natural
  Computation. Volume~1. (2007)  283--286

\bibitem{YaoVasilakosPedrycz2013Granular}
Yao, J.T., Vasilakos, A.V., Pedrycz, W.:
\newblock Granular computing: Perspectives and challenges.
\newblock IEEE Transactions on Systems, Man, and Cybernetics, Part C:
  Applications and Reviews \textbf{PP}(99) (2013)  1--13

\bibitem{YaoDeng2013Paradigm}
Yao, Y.Y., Deng, X.F.:
\newblock A granular computing paradigm for concept learning.
\newblock In: Emerging Paradigms in Machine Learning. Volume~13.
\newblock Springer Berlin Heidelberg (2013)  307--326

\bibitem{MinHeQianZhu11Test}
Min, F., He, H.P., Qian, Y.H., Zhu, W.:
\newblock Test-cost-sensitive attribute reduction.
\newblock Information Sciences \textbf{181} (2011)  4928--4942

\bibitem{movielens}
Internet movie database, http://movielens.umn.edu

\bibitem{Grale}
Min, F., Zhu, W., He, X.:
\newblock Grale: Granular association rules,
  http://grc.fjzs.edu.cn/\~{}fmin/grale/ (2013)

\bibitem{Pawlak82Rough}
Pawlak, Z.:
\newblock Rough sets.
\newblock International Journal of Computer and Information Sciences
  \textbf{11} (1982)  341--356

\bibitem{SkowronStepaniuk94Approximation}
Skowron, A., Stepaniuk, J.:
\newblock Approximation of relations.
\newblock In Ziarko, W., ed.: Proceedings of Rough Sets, Fuzzy Sets and
  Knowledge Discovery. (1994)  161--166

\end{thebibliography}

\end{document}